# Temperature Dependence Cancellation of the Cs Clock Frequency in the Presence of Ne Buffer Gas

Olga Kozlova, Rodolphe Boudot[1], Stéphane Guérandel, Emeric de Clercq
Laboratoire National d'Essai-Systèmes de Référence Temps Espace
Observatoire de Paris, UMR 8630, CNRS, Université Pierre et Marie Curie, Paris, France

***Abstract*—The temperature dependence of the Cs clock transition frequency in a vapor cell filled with Ne buffer gas has been measured. The experimental setup is based on the coherent population trapping (CPT) technique and a temporal Ramsey interrogation allowing a high resolution. A quadratic dependence of the frequency shift is shown. The temperature of the shift cancellation is evaluated. The actual Ne pressure in the cell is determined from the frequency shift of the 895nm optical transition. We can then determine the Cs-Ne collisional temperature coefficients of the clock frequency. These results can be useful for vapor cell clocks and especially for future micro-clocks.**

## I. INTRODUCTION

THE coherent population trapping (CPT) phenomenon [1] has been investigated by several laboratories for a few years in view of applications to atomic clocks [2, 3, 4]. The CPT has renewed the interest in the use of the cesium atom in atomic vapour cell clocks, when they were until now based on rubidium. In such clocks a buffer gas is added in the alkali vapor cell in order to increase the interaction time by reducing the effect of wall collisions and in order to reduce the Doppler width by the Dicke effect. However the collisions between alkali atoms and the buffer gas can shift the frequency of the resonance by several kHz per Torr, and this shift (known as collisional or pressure shift) is temperature dependent.

Over a limited range of temperatures it is generally possible to represent the temperature dependence of the collisional shift $\Delta\upsilon(T)$ by a quadratic equation [5]:

$$\Delta\upsilon(T) = P_0\,[\,\beta + \delta(T-T_0) + \gamma(T-T_0)^2], \qquad (1)$$

where $P_0$ is the buffer gas pressure in the cell at the reference temperature $T_0$, $\beta$ is the pressure coefficient ($Hz\cdot Torr^{-1}$), $\delta$ is the linear temperature coefficient ($Hz\cdot Torr^{-1}\cdot K^{-1}$), $\gamma$ is the quadratic temperature coefficient ($Hz\cdot Torr^{-1}\cdot K^{-2}$) and $T$ is the buffer gas temperature.

Using a mixture of gases having temperature coefficients with opposite signs, one can cancel the temperature dependence at the working temperature. Another possibility to cancel the temperature dependence using only one buffer gas is to work near the inversion temperature (for buffer gases exhibiting strong quadratic temperature dependence). This has already been measured for Rb atoms [6, 7] but to our knowledge not for Cs atoms.

In this paper we report the measurement of the temperature coefficients of the Cs clock frequency [8] in the presence of Ne buffer gas. To determine the value of the coefficient we have measured the shift of the clock frequency against the cell temperature. The combination of two original techniques [9]: a double-lambda scheme for the CPT-resonance excitation and a temporal Ramsey interrogation technique allows to obtain a high contrast and narrow resonances with the width scaling as $1/[2T_R]$, where $T_R$ is the time separation between the light pulses. It has been shown in the pulsed (Ramsey) regime that the CPT linewidth is not limited by the saturation effect and light shift dependence is greatly reduced [10]. This method provides a better frequency uncertainty that previously reported ones in Cs (Optical Pumping [11] or CPT [12] methods) and allows to reveal the quadratic shift dependence.

The inversion temperature does not depend on the gas pressure and can be directly obtained from the temperature dependence. On the contrary the knowledge of the actual buffer gas pressure in the cell is necessary to establish the values of the coefficients. It is the major source of uncertainty. We have measured it from the shift of the optical transitions ($D_1$ line).

In this paper, Section II is devoted to the measurement of the Cs hyperfine frequency shift as a function of the temperature of sealed cells filled with Ne. The experimental setup is first described. Experimental results are shown, the quadratic dependence is clearly visible. The importance of the inversion temperature for Ne buffer gas for applications in chip scale atomic clocks is discussed. In Section III, the measurement of the buffer gas pressure in the sealed cells is presented. In Section IV, the measurement results of the pressure coefficient $\beta$ ($Hz\cdot Torr^{-1}$), linear temperature coefficient $\delta$ ($Hz\cdot Torr^{-1}\cdot K^{-1}$) and quadratic temperature coefficient $\gamma$ ($Hz\cdot Torr^{-1}\cdot K^{-2}$) for Ne buffer gas are reported.

[1] R. Boudot is now with Franche-Comté Electronique, Mécanique, Thermique et Optique-Sciences et Technologies (FEMTO-ST) Institute, Unité Mixte de Recherche 6174, CNRS and Université de Franche Comté, Besançon, France.

## II. Measurements of the Temperature Dependence of the Cs Clock Transition in the Presence of Ne Buffer Gas

### A. *Experimental setup*

The experimental setup used for temperature shift measurements is shown on Fig. 1. The optical radiations required to pump the atoms into the CPT state (which is a superposition state of the clock levels |$F$=3, $m$=0> and |$F$=4, $m$=0>) are generated by two external cavity diode lasers (ECDL) tuned to the Cs $D_1$ line at 895 nm. The master laser is frequency locked to the $F = 4 - F' = 4$ hyperfine component of the Cs D1 line by a saturated absorption scheme in an auxiliary pure Cs vapor cell. The slave laser is phased locked to the master laser with a frequency offset that is tunable near 9.192 GHz by comparison with a low noise synthesized microwave signal [13] driven by a hydrogen maser. For this purpose, the two laser beams are superimposed and detected by the fast photodiode PD2.

We use a so-called double-lambda scheme with two linear and orthogonally polarized laser beams (lin perp lin configuration) propagating parallel to the static magnetic field applied to the cell. The atomic response is detected by measuring the optical power of the laser beams transmitted through the cell.

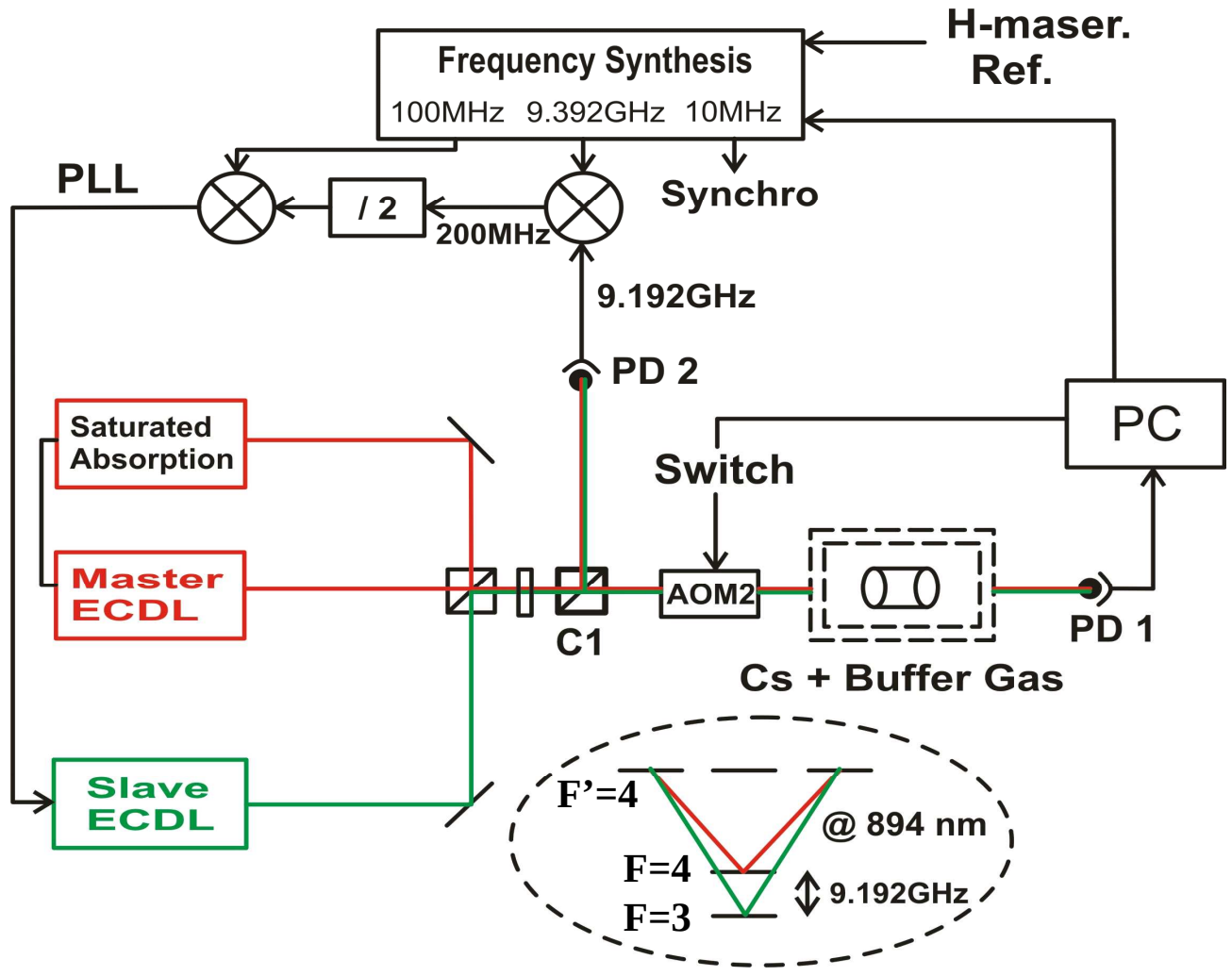

Fig. 1. Experimental setup for temperature shift measurements. AOM: acousto-optical modulator, PD1: atomic signal photodetector, PD2: fast silicon photodiode, ECDL: external cavity diode laser. The inset shows the atomic energy levels involved in the double-lambda scheme. PLL: phase lock loop. C1: beam splitter cube.

A CPT pulse sequence where each pulse is used both for coherence preparation and atomic signal detection is applied. The light pulse durations ($\tau$ = 2 ms) and the time separation between the pulses ($T_R$ = 4 ms) are controlled by an acousto-optic modulator. There is a possibility to work in CW operation mode by applying a constant power RF level to the acousto-optic modulator.

We use two sealed Cs vapor cells filled with 90 Torr Ne as buffer gas made by two different providers: cell A (20 mm long and 20 mm diameter) from Physics Department, Fribourg University, Switzerland, and the cell B (50 mm long and diameter 25 mm) from Triad company, USA. The temperature of the cell is measured by high accuracy nonmagnetic thermistors with a 0.2°C uncertainty. It can be changed in the 25°C - 65°C temperature range and stabilized within the mK level. The temperature can be risen up to 80°C and stabilized at the level of tens of mK. A static magnetic field is applied with a solenoid to lift the degeneracy of the Zeeman levels. The ensemble is surrounded by two mu-metal magnetic shields.

For each Cs-buffer-gas cell the resonance frequency is measured for a series of temperatures. For each measurement point, the cell temperature is allowed at least 2 hours to stabilize. Each measurement is performed by locking the 9.2 GHz synthesizer frequency on the central Ramsey fringe. For this purpose the synthesizer frequency is square wave modulated. The absorption signal (PD1) is digitized, computer-processed and the correction signal is applied to the synthesizer. The mean frequency is averaged during a few hours. It is referenced to the hydrogen maser, whose frequency is compared to the Cs clocks of the laboratory.

The measured frequencies are corrected for the Zeeman quadratic shift and for the light shift. For this purpose, the frequency is measured at different laser power level for each temperature. The frequency value is then extrapolated to the zero power.

### B. *Experimental results*

The temperature dependence of the collisional frequency shift is shown on Fig. 2 and Fig. 3 for the cell A and B respectively. The frequency shift is measured using the Ramsey technique in the 25°C - 65°C temperature range. The frequency is measured with a 0.1 Hz and 0.3 - 0.5 Hz accuracy in the 25°C - 45°C and 45°C - 65°C temperature ranges respectively.

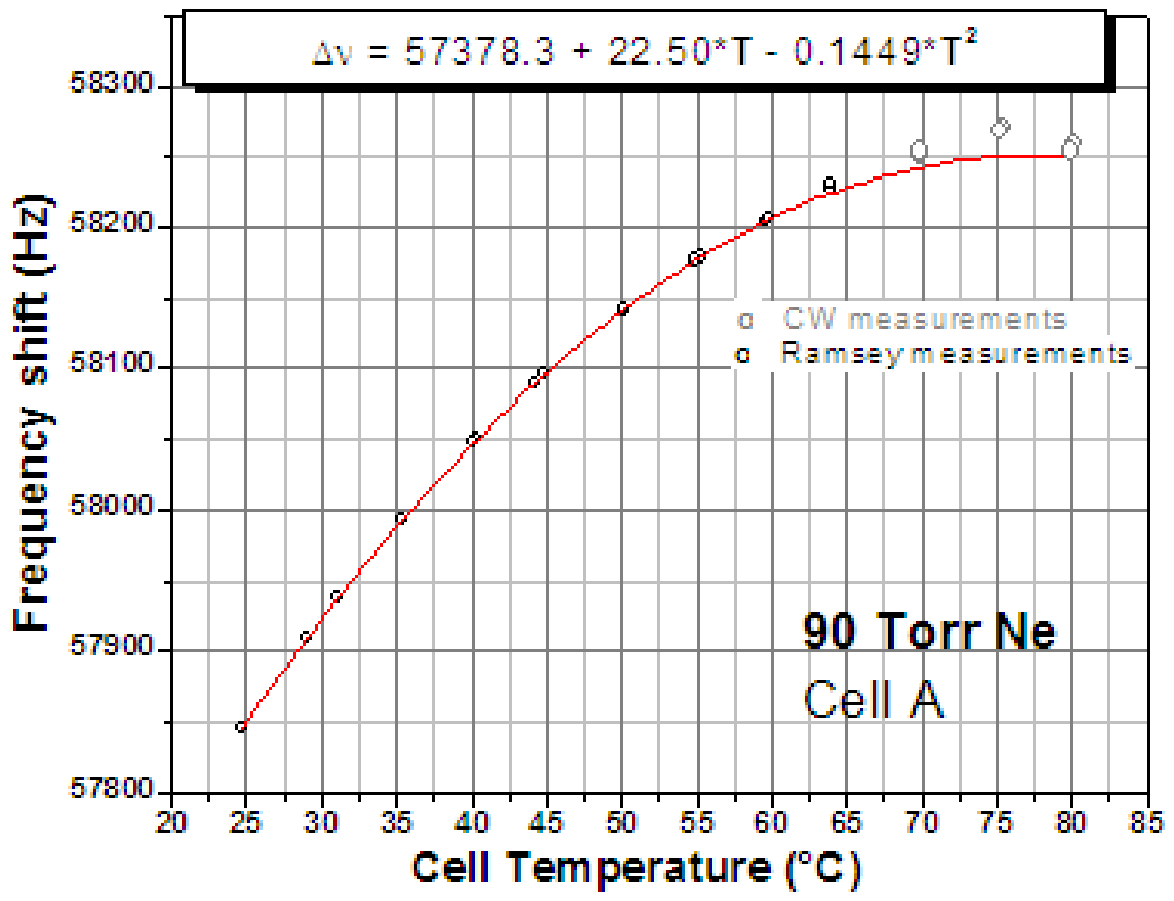

Fig. 2. Cell A - Temperature dependence of the collisional shift of the Cs microwave clock transition in the presence of Ne buffer gas. Black circles: measurements with Ramsey method, grey circles: continuous method. The error bars are within the circles. The equation of the fitted curve is given in the inset, $T$ is the cell temperature in °C.

The experimental points are fitted with a 2$^{nd}$ order polynomial function as reported in Eq. 1. Then from the fitting coefficients, we find that the Cs clock frequency temperature dependence is cancelled in the 75°C - 85°C temperature range. For cell A, the measurement of the clock frequency versus temperature was made in the 70°C - 80°C range in order to confirm the inversion temperature. However, in this temperature range, the increase of the optical absorption in the vapor cell and of the number of alkali-buffer gas and Cs-Cs spin-exchange collisions causes a significant reduction of the coherence lifetime and of the Ramsey fringes amplitude. Then, the Ramsey interrogation technique becomes difficult to apply. Consequently, the measurements are performed in this temperature range using a classical continuous interrogation technique (instead of the pulsed Ramsey technique) at the expense of a reduced accuracy (1-10Hz) (see Fig. 2).

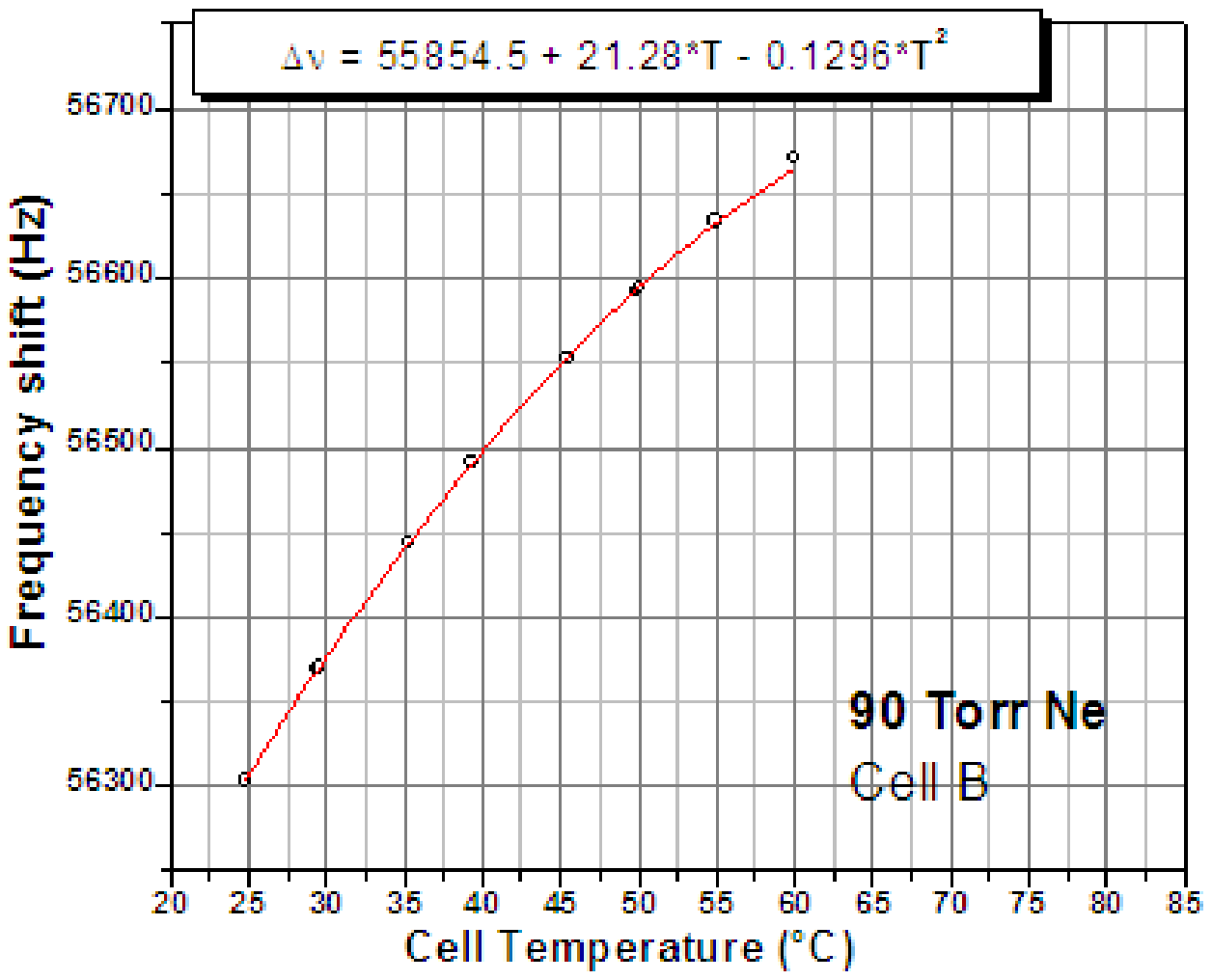

Fig. 3. Cell B - Temperature dependence of the collisional shift of the Cs microwave clock transition in Cs in the presence of Ne buffer gas. The error bars are within the circles. The equation of the fitted curve is given in the inset, $T$ is the cell temperature in °C.

The inversion temperature does not depend on the pressure of the buffer gas, so no pressure measurement is needed. The results for the inversion temperature for Cs in presence of Ne buffer gas are reported in Table I.

The lack of precision on the value of the inversion temperature is due to the low precision on the high temperature data and lack of data points above the inversion temperature, which reduce the accuracy of the fit coefficients. We plan to realize higher temperature measurements with shorter cells to overcome these difficulties. With a conservative point of view we can determine the inversion temperature $T_i$ as the weighted mean of our values, and their difference as uncertainty. This yields to $T_i = (80 \pm 4)$°C.

TABLE I
INVERSION TEMPERATURE FOR CS MICROWAVE CLOCK TRANSITION IN PRESENCE OF NE BUFFER GAS

| Cs Cells used | Inversion temperature $T_i$ | Uncertainty | Reference |
|---|---|---|---|
| (A) 90 Torr Ne | 77.7°C | 5°C | This work |
| (B) 90 Torr Ne | 82.0°C | 7°C | This work |
| 75 Torr Ne | 80°C | N.A. | [14] |
| 2028 Torr Ne | 83°C | N.A. | from a figure of [15] |

Recently [14], the inversion temperature was observed in millimeter size Cs cells with a 75 Torr pressure of Ne buffer gas. The results are in agreement with ours. In [15] the quadratic temperature dependence of the density shift is visible on a figure, but no details on the coefficients or the inversion temperature are given. We inferred the inversion temperature from the figure published in [15].

It is important to notice, that the inversion temperature for Cs in Ne buffer gas is in the working range of Cs chip scale CPT clocks. It means that no mixture of buffer gas is required for development of such a clock, simplifying potentially the production process and reducing the cost of the device.

## III. Measurements of the Buffer Gas Pressure from the Shift of the Optical Transitions

### A. *Experimental setup*

The usual technique to cancel (or reduce) the temperature dependence (at a chosen working temperature) is to fill the cell with a mixture of two buffer gases which shift the clock frequency in opposite directions. The inversion temperature can be chosen by fixing the pressure ratio of the two buffer gases. This needs the precise knowledge of the temperature coefficients ($\beta$, $\delta$, and $\gamma$) of each buffer gas. In order to measure the coefficients of Ne the actual Ne pressure in the cell is needed because its value can be altered during the high temperature sealing process. We measure its pressure by measuring the shift of the absorption spectrum of the Cs optical transition (D1 line).

The experimental setup for the measurement of the optical shift in Cs cells with buffer gas is shown on Fig. 4.

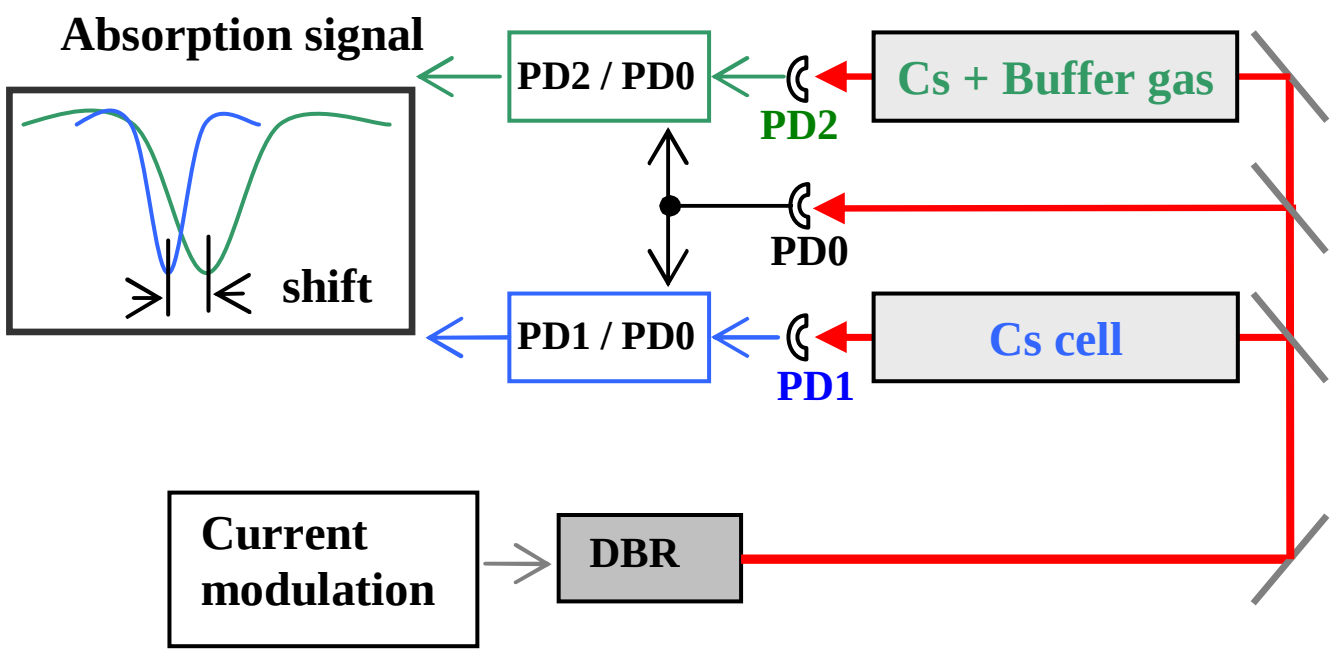

Fig. 4. Experimental setup for optical shift measurements. DBR - Distributed Bragg Reflector laser diode; PD0, PD1 and PD2 – photodetectors.

We use a Distributed Bragg Reflector laser diode (DBR) tuned to Cs $D_1$ line. A slow current modulation is applied in order to scan the frequency over 12 GHz. We work with very low intensities (less than 10 μW/cm$^2$) to avoid optical pumping effects. The power transmitted through the pure Cs cell and the Cs+buffer gas cell is detected by photodiodes PD1 and PD2, respectively. The photodiode PD0 is used to register the power changes during the frequency scanning. Then the power changes are eliminated from the PD1 and PD2 signals by normalization.

The absorption signals from the Cs cell and the Cs+buffer gas cell are registered simultaneously. As the resonances are significantly broadened and overlapped (in the case of the Cs+buffer gas cells) the line centers are determined by a numerical fit of the spectrum. We use a sum of Voigt profiles to fit the absorption spectrum (Fig.5). The positions of the peaks are extracted from the fit.

To calibrate the *x*-axis (frequency axis) we correlate the difference in the peaks position of the pure Cs (without buffer gas) absorption curve with the well-known hyperfine frequency structure of D1 line (see inset in Fig.5).

The mean shift of the peaks position in the Cs+buffer gas cell compared to the pure Cs cell is the optical shift proportional to the buffer gas pressure.

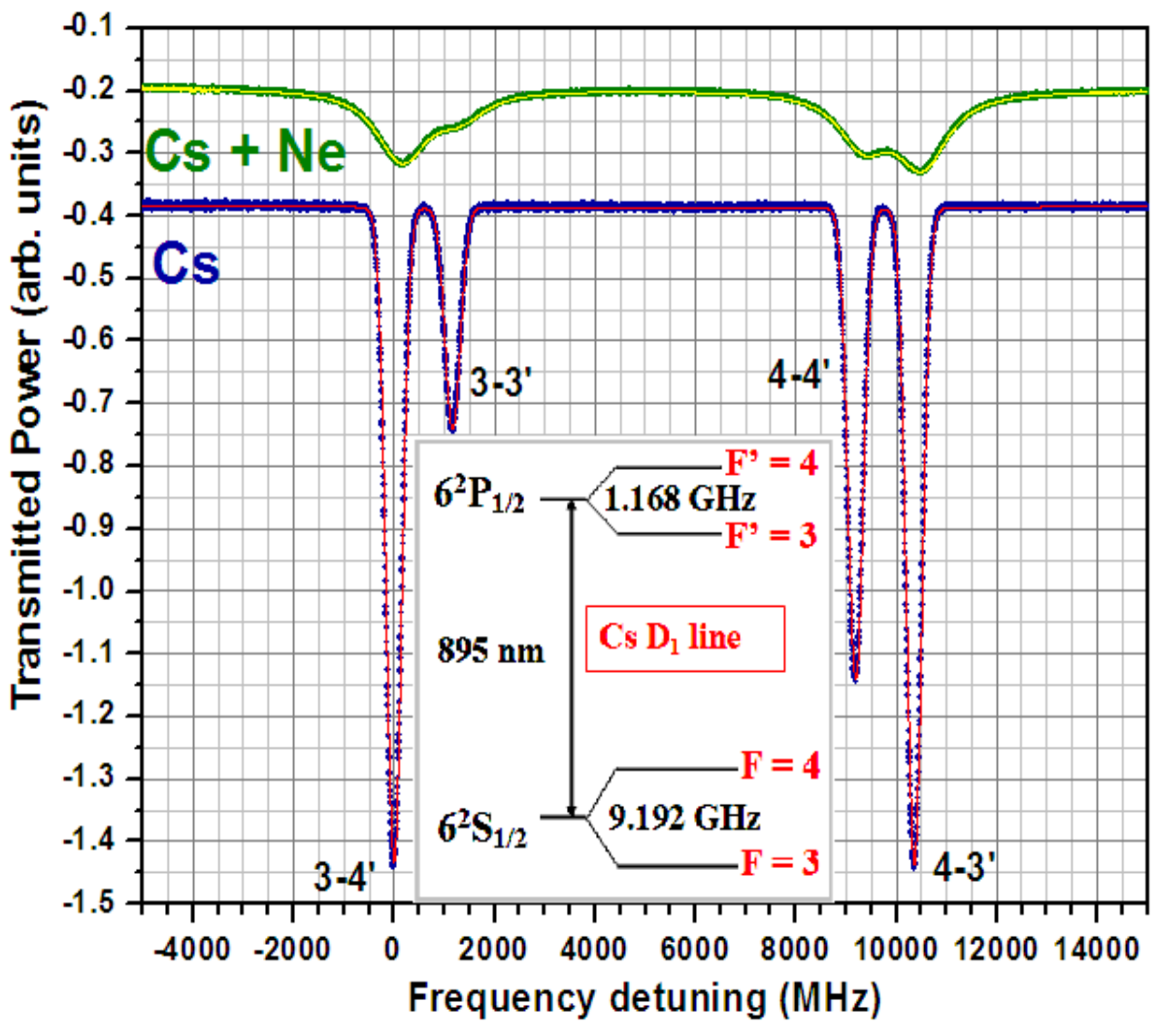

Fig. 5. Absorption spectra. Dark blue and green dots – experimental curves for Cs and Cs+buffer gas cell, respectively. Red and yellow lines – fit with Voigt profile for Cs and Cs+buffer gas cells, respectively. The inset shows the involved energy levels.

### B. *Buffer gas pressure results*

To determine the actual buffer gas pressure in the cell from the measured values of the shift of the optical transitions, we use the pressure shift coefficients for optical transitions published by Pitz *et al.* [16]. The optical shift values for two buffer gas cells (cell A, cell B) and calculated values of the buffer gas pressure in the cells are presented in Table II.

TABLE II
OPTICAL SHIFT AND PRESSURE VALUES FOR NE BUFFER GAS

| Cell | Optical shift value at 22°C (MHz) | Pressure shift coefficient [16] (MHz/torr) | Pressure value at 22°C (Torr) | Pressure value at 0°C (Torr) |
|---|---|---|---|---|
| 90 Torr Ne (cell A) | - 154.0 ± 3 | -1.60 ± 0.01 | 96.3 ± 3 | 89.1 ± 3 |
| 90 Torr Ne (cell B) | - 143.6 ± 3 | -1.60 ± 0.01 | 89.8 ± 3 | 83.1 ± 3 |

## IV. RESULTS FOR PRESSURE AND TEMPERATURE COEFFICIENTS FOR CS CLOCK TRANSITION

Knowing the gas pressure in the cells we can extract the hyperfine transition coefficients values from the fitted curves in Fig. 2 and Fig. 3. The measured pressure and temperature coefficients for Cs clock transition are shown in Table III and compared with previously published values when they are available. For the first time the quadratic coefficients for Ne buffer gas is determined. The measured pressure coefficient $\beta$ is in good agreement with previous measurements [15, 17]. The agreement between the values of the linear temperature coefficient $\delta$ is not so good. Our value is about twice the value given by Beverini *et al* [17]. However the discrepancy is slightly larger than 1$\sigma$ of [17]. There is no possible comparison for the quadratic coefficient $\gamma$ which is measured for the first time. It is worthwhile to note that the two cells, denoted A and B, are issued from two independent manufacturers, each one using its own process. The good agreement between results obtained on the two cells shows that we can be confident in the results. We can estimate that no significant bias shift due to the cell production is present.

Note that the inversion temperature computed from the coefficients mean value is *Ti* = (80 ± 6)°C. The uncertainty is larger than the one given in Section II because the uncertainty on the Ne pressure enters in the uncertainties of the coefficients.

TABLE III
PRESSURE COEFFICIENT, LINEAR AND QUADRATIC TEMPERATURE COEFFICIENTS FOR CS IN PRESENCE OF NE BUFFER GAS

| pressure coefficient β (Hz/Torr) | linear temperature coefficient δ (Hz/Torr K) | quadratic temperature coefficient γ (Hz/Torr $K^2$) | Reference |
|---|---|---|---|
| 644.0 ± 32 | 0.253 ± 0.013 | (-1.62 ± 0.09) x $10^{-3}$ | This work (Cell A) |
| 672.1 ± 38 | 0.256 ± 0.015 | (-1.56 ± 0.09) x $10^{-3}$ | This work (Cell B) |
| 657 ± 25 | 0.254 ± 0.010 | (-1.59 ± 0.06) x $10^{-3}$ | This work (weighted mean) |
| 647 ± 4 | - | - | [15] |
| 652 ± 20 | 0.14 ± 0.10 | - | [17] |

## V. CONCLUSION

We have measured the dependence of the Cs clock frequency in two Cs vapor cells filled with about 90 torr of Ne. The quadratic behavior of the frequency is clearly pointed out. The temperature shift vanishes around 80°C. This result is promising for future miniature atomic clocks using mm-scale cell. It allows the use of a single buffer gas, and the inversion temperature is in their range of working temperature.

In order to calculate the Cs-Ne collisional shift coefficients we have measured the Ne pressure in the sealed cells. For this purpose, the frequency shift of the Cs D1 line is measured, and then the buffer gas pressure is deduced from the shift using pressure shift coefficients published in [17]. We report the values of the collisional coefficients. To our knowledge it is the first measurement of the quadratic temperature term. These coefficients are needed to foresee the compound of a buffer gases mixture allowing a cancellation of the temperature shift at an arbitrary chosen temperature.

## ACKNOWLEDGMENT

We are grateful to Michel Lours for his helpful assistance. We also thank Laurent Volodimer and José Pinto for their contributions to the electronic parts of the experiments. We thank Ouali Acef for his support. We are pleased to acknowledge G. P. Perram, G. A. Pitz, and Peter Wolf for valuable advices on the fitting method. We thank Yann le Coq for attentive reading of the manuscript.
The work of O. Kozlova was supported by DGA. This work was supported by DGA, contract REI N° 2009.34.0052.00000000.